\documentclass[%
reprint,
superscriptaddress,
 amsmath,amssymb,
prl, 
aps,
]{revtex4-2}

\usepackage{graphicx}
\usepackage{dcolumn}
\usepackage{bm}
\usepackage[detect-all]{siunitx}

\begin{document}

\title{Efficient Acceleration of High-Quality GeV-Electron Bunches in a Hybrid Laser- and Beam-Driven Plasma Wakefield Accelerator}

\author{F. M. Foerster}
\email{moritz.foerster@physik.uni-muenchen.de}
\affiliation{Ludwig--Maximilians--Universit{\"a}t M{\"u}nchen, Am Coulombwall 1, 85748 Garching, Germany}%

\author{M. Ayache}
\affiliation{Ludwig--Maximilians--Universit{\"a}t M{\"u}nchen, Am Coulombwall 1, 85748 Garching, Germany}%

\author{Z. Bi}
\affiliation{Ludwig--Maximilians--Universit{\"a}t M{\"u}nchen, Am Coulombwall 1, 85748 Garching, Germany}%

\author{M. Cerchez}
\affiliation{Heinrich Heine Universität Düsseldorf, Universitätsstraße 1, 40225 Düsseldorf, Germany}%

\author{S. Corde}
\affiliation{Laboratoire d’Optique Appliquée (LOA), CNRS, École polytechnique, ENSTA, Institut Polytechnique de Paris, Palaiseau, France}%

\author{A. D{\"o}pp}
\affiliation{Ludwig--Maximilians--Universit{\"a}t M{\"u}nchen, Am Coulombwall 1, 85748 Garching, Germany}%
\affiliation{Max Planck Institut für Quantenoptik, Hans-Kopfermann-Strasse 1, 85748 Garching , Germany}%

\author{F. Haberstroh}
\affiliation{Ludwig--Maximilians--Universit{\"a}t M{\"u}nchen, Am Coulombwall 1, 85748 Garching, Germany}%

\author{A. F. Habib}
\affiliation{University of Strathclyde, 107 Rottenrow, Glasgow G4 0NG, United Kingdom}

\author{T. Heinemann}
\affiliation{Heinrich Heine Universität Düsseldorf, Universitätsstraße 1, 40225 Düsseldorf, Germany}%

\author{B. Hidding}
\affiliation{Heinrich Heine Universität Düsseldorf, Universitätsstraße 1, 40225 Düsseldorf, Germany}%

\author{A. Irman}
\affiliation{Helmholtz-Zentrum Dresden--Rossendorf, Bautzner Landstrasse 400, 01328 Dresden, Germany}%

\author{F. Irshad}
\affiliation{Ludwig--Maximilians--Universit{\"a}t M{\"u}nchen, Am Coulombwall 1, 85748 Garching, Germany}%

\author{O. Kononenko}
\affiliation{Laboratoire d’Optique Appliquée (LOA), CNRS, École polytechnique, ENSTA, Institut Polytechnique de Paris, Palaiseau, France}%

\author{M. LaBerge}
\affiliation{Helmholtz-Zentrum Dresden--Rossendorf, Bautzner Landstrasse 400, 01328 Dresden, Germany}%

\author{A. Martinez de la Ossa}
\affiliation{Deutsches Elektronen-Synchrotron DESY, Notkestraße 85, 22607 Hamburg, Germany}%

\author{A. Münzer}
\affiliation{Ludwig--Maximilians--Universit{\"a}t M{\"u}nchen, Am Coulombwall 1, 85748 Garching, Germany}%

\author{F. Pe\~na}
\affiliation{Ludwig--Maximilians--Universit{\"a}t M{\"u}nchen, Am Coulombwall 1, 85748 Garching, Germany}%
\affiliation{Department of Physics, University of Oslo, 0316, Oslo, Norway}

\author{G. Schilling}
\affiliation{Ludwig--Maximilians--Universit{\"a}t M{\"u}nchen, Am Coulombwall 1, 85748 Garching, Germany}%

\author{S. Sch{\"o}bel}
\affiliation{Helmholtz-Zentrum Dresden--Rossendorf, Bautzner Landstrasse 400, 01328 Dresden, Germany}%
\affiliation{Technische Universität Dresden, 01062 Dresden, Germany}

\author{U. Schramm}
\affiliation{Helmholtz-Zentrum Dresden--Rossendorf, Bautzner Landstrasse 400, 01328 Dresden, Germany}
\affiliation{Technische Universität Dresden, 01062 Dresden, Germany}%

\author{S. Sharan}
\affiliation{Ludwig--Maximilians--Universit{\"a}t M{\"u}nchen, Am Coulombwall 1, 85748 Garching, Germany}%

\author{E. Travac}
\affiliation{Ludwig--Maximilians--Universit{\"a}t M{\"u}nchen, Am Coulombwall 1, 85748 Garching, Germany}%

\author{P. Ufer}
\affiliation{Helmholtz-Zentrum Dresden--Rossendorf, Bautzner Landstrasse 400, 01328 Dresden, Germany}%
\affiliation{Technische Universität Dresden, 01062 Dresden, Germany}

\author{N. Weiße}
\affiliation{Ludwig--Maximilians--Universit{\"a}t M{\"u}nchen, Am Coulombwall 1, 85748 Garching, Germany}%

\author{M. Zeuner}
\affiliation{Ludwig--Maximilians--Universit{\"a}t M{\"u}nchen, Am Coulombwall 1, 85748 Garching, Germany}%

\author{J. Zirkelbach}
\affiliation{Ludwig--Maximilians--Universit{\"a}t M{\"u}nchen, Am Coulombwall 1, 85748 Garching, Germany}%

\author{S. Karsch}
\email{stefan.karsch@physik.uni-muenchen.de}
\affiliation{Ludwig--Maximilians--Universit{\"a}t M{\"u}nchen, Am Coulombwall 1, 85748 Garching, Germany}%
\affiliation{Max Planck Institut für Quantenoptik, Hans-Kopfermann-Strasse 1, 85748 Garching , Germany}%

\date{\today}

\begin{abstract}
Plasma-based accelerators are compact and provide high gradients, yet their practical use has been limited by energy gain, stability, beam quality, and energy transfer efficiency. 
Here, we address several of these challenges simultaneously using a hybrid scheme in which an electron bunch from a laser wakefield accelerator (LWFA) drives a subsequent plasma wakefield accelerator (PWFA) stage with internal witness injection. 
Close to driver depletion in the PWFA stage, we obtain witness bunches with higher electron energy, reduced energy spread and divergence, and higher angular-spectral charge density compared to LWFA alone. 
We report energy transformer ratios approaching~2, and about 20\% of the initial energy in the drive beam was transferred to the witness bunch, thereby achieving a driver-to-witness energy transfer efficiency that largely surpasses that of all previous PWFA experiments.
\end{abstract}

\maketitle


\emph{Introduction.}--The demand for more compact and cost-effective particle accelerators has spurred the exploration of novel acceleration concepts beyond conventional radio-frequency (rf) technologies. Among these, plasma-based accelerators have emerged as a promising candidate due to their ability to sustain accelerating gradients several orders of magnitude higher than those in rf-accelerating structures~\cite{tajima1979, esarey2009,lindstrom2025}.  
Laser- and beam-driven plasma wakefield accelerators (LWFAs and PWFAs) have demonstrated significant progress in recent years, providing electron bunches with ultrashort duration~\cite{Heigoldt:2015cd,zarini2022,laberge2024} and high peak currents~\cite{Gotzfried:2020da,Couperus:2017bg,emma2025}. Over the years, plasma accelerators have seen steady improvements in energy gain~\cite{Blumenfeld:2007ja,Gonsalves:2019ht,picksley2024,shrock2024,zhang2025}, spectral quality~\cite{Lindstrom:2021eo,Ke:2021hh}, stability~\cite{foerster2022}, and energy transfer efficiency~\cite{Litos:2016,Gotzfried:2020da}. However, optimizing multiple beam parameters simultaneously remains a major challenge, as trade-offs between these often limit performance in practical scenarios. Application-oriented beams demand a careful balance between these objectives, motivating new approaches that can break current limitations.

A promising path forward is offered by the hybrid laser- and beam-driven plasma wakefield accelerator (LWFA-PWFA), which aims to combine the complementary strengths of LWFA and PWFA stages~\cite{hidding2023,Chou:2016ei,Gilljohann:2019kc,Kurz:2021cg, Gotzfried:2020da,CouperusCabadag:2021gj,schobel2022,foerster2022}. LWFAs benefit from the wide availability of high-power laser systems. However, their operation is challenged by the combined effects of dephasing between the electrons and the driving laser pulse, laser diffraction, and plasma-electron heating from the laser’s oscillatory field. Although each of these issues can be mitigated individually, their simultaneous presence makes it difficult to achieve high-energy and ultralow-emittance electron bunches. 
Moreover, the properties of electron bunches obtained from LWFA are particularly susceptible to shot-to-shot fluctuations in the laser output and its subsequent nonlinear propagation in the plasma~\cite{Maier:2020kh,oumbarekespinos2023}. 
In contrast, PWFAs using ultrarelativistic charged particle beams as drivers are dephasingless and, when combined with suitable injection schemes, could produce witness bunches with exceptionally low transverse emittance~\cite{MartinezdelaOssa:2015es,Hidding:2012ep}. Their main limitation lies in the scarcity of suitable drive bunches, which typically require access to large-scale accelerator facilities~\cite{author2016,darcy2019,Adli:2018jza,pompili2021}. By leveraging an LWFA-generated electron bunch to drive a subsequent PWFA stage, the hybrid approach offers a practical route to harness the advantages of both plasma accelerator concepts within a compact setup.

\begin{figure*}[t]
  \centering
  \includegraphics*[width=\linewidth]{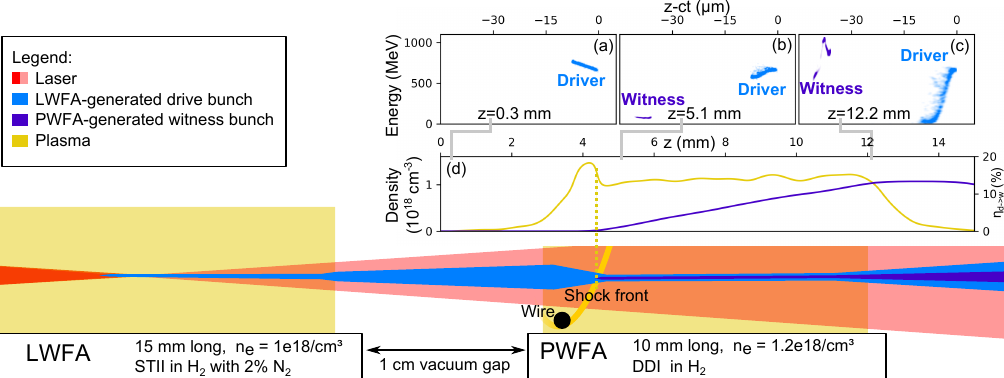}
  \caption{Schematic setup for the experimental generation of high-energy, high-quality witness bunches in a hybrid LWFA-PWFA. In the LWFA stage, electron bunches are generated via self-truncated ionization injection (STII). After a vacuum gap, these electron bunches (light blue) drive the PWFA stage. A wire-generated shock (orange) enables the density down ramp injection (DDI) of witness electrons (violet) into the PWFA stage. Insets (a)-(c) show the longitudinal phase space of driver and witness retrieved from an exemplary PIC simulation at different positions in the PWFA stage~\cite{Note1}. (d) shows the plasma density distribution of the PWFA together with the evolution of the driver-to-witness energy transfer efficiency along the length of the PWFA stage.}
  \label{fig:setup}
\end{figure*}

The hybrid LWFA-PWFA concept evolved from the observation of collective deceleration of LWFA-generated electrons in a plasma~\cite{Chou:2016ei,Gilljohann:2019kc}, to a dedicated two-stage configuration, although with limited control over witness injection in the PWFA stage~\cite{Kurz:2021cg}, and eventually to controlled internal down ramp injection of witness bunches in the PWFA stage~\cite{CouperusCabadag:2021gj}. More recently, we have begun leveraging the full potential of this configuration, demonstrating high stability and an enhanced angular–spectral charge density of the witness bunch compared with the initial LWFA-generated drive bunch~\cite{foerster2022}. While these experiments have revealed the potential of hybrid LWFA-PWFA as a beam quality booster~\cite{MartinezdelaOssa:2019bm}, a booster for the peak electron energy~\cite{hidding2010} has not yet been experimentally verified. In all previous experiments, the peak electron energy of the witness bunch has remained limited to values lower than that of the driver, and the driver-to-witness energy transfer efficiency of the PWFA stage has not yet been fully exploited.

In this work, we present witness bunches from a hybrid LWFA-PWFA setup, characterized by higher electron energy, lower energy spread, and lower divergence than the LWFA-generated drive bunches. At the same time, the driver-to-witness energy transfer efficiency reaches unprecedentedly high values of $\sim20\%$. Furthermore, we conduct a careful review of various measures used in the literature to characterize the electron energy gain and efficiency of plasma-based accelerators. Based on this, we compare our results to the current state of the art.

\emph{Experimental setup.}--The experiments were performed at the Centre for Advanced Laser Applications in Garching, Germany. The hybrid LWFA-PWFA setup consists of two consecutive gas targets separated by a 1 cm vacuum gap (see Fig.~\ref{fig:setup}). The first stage is an LWFA, driven by a Ti:sapphire laser pulse with $\approx\SI{9}{J}$ on-target energy, $\SI{30} {fs}$ (FWHM) pulse duration at a central wavelength of $\SI{800} {nm}$. The laser pulse is focused with an f/47 geometry to a peak intensity of $(1.6\pm0.3)\times10^{19}\:\si{\watt\per\cm^2}$, which corresponds to a normalized vector potential of $a_0\approx 2.7$.
The target is a 15 mm-long slit nozzle operated with hydrogen gas doped with 2\% nitrogen gas at a plasma density of $\SI{1.0e18}{cm^{-3}}$. Electron injection in the LWFA occurs via self-truncated ionization injection~\cite{Couperus:2017bg,irman2018,mirzaie2015,zeng2014}, which relies on rapid laser self-focusing in a low-ionization-threshold background gas, followed by ionization of inner-shell electrons from the dopant species. This configuration produces electron bunches with charge of $366\pm69\,\si{pC}$ (mean and standard deviation of the integrated charge above $\SI{200}{MeV}$), a rms-divergence 
of $0.9\pm0.1\,\si{mrad}$ (mean and standard deviation from gaussian fit), a peak electron energy of $716\pm44\,\si{MeV}$ (mean and standard deviation), and a relative FWHM-energy spread of $9\%\pm3\%$ (mean and standard deviation) as measured using an absolutely calibrated dipole spectrometer. 

The subsequent PWFA stage employs the LWFA-generated electron bunch as the driver. The beam axis was positioned close ($\SI{2}{mm}$) above the exit face of the 10 mm-long supersonic slit nozzle to minimize density ramps at both ends (see Fig.~\ref{fig:setup}(d)). The remaining laser from the LWFA stage ionizes the gas homogeneously in a $\sim \SI{1}{mm}$ wide column around the optical axis. However, it is no longer intense enough to excite a wakefield itself in the second target after diffracting in the 1-cm vacuum gap between the stages~\cite{foerster2022}. The PWFA stage is operated with pure hydrogen gas at a plasma density of $\SI{1.2e18}{cm^{-3}}$. 
Controlled injection of a witness beam is achieved via a hydrodynamic shock, generated by placing a thin wire in the supersonic gas flow. The resulting localized density transition provides conditions suitable for density down ramp injection into the wakefield~\cite{CouperusCabadag:2021gj,foerster2022}.

\begin{figure}[t]
  \centering
  \includegraphics*[width=\linewidth]{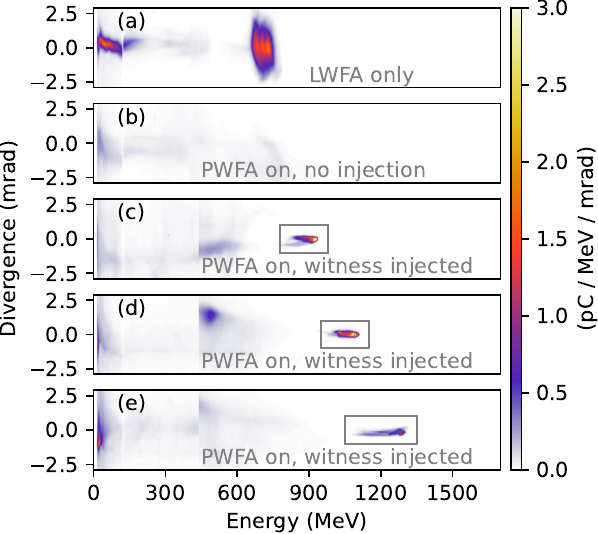}
  \caption{Energy spectra of electron driver and witness for various conditions of hybrid LWFA-PWFA. (a) LWFA-generated drive beam. (b) Decelerated drive beam after the PWFA stage, without internal injection. (c) Decelerated driver and a typical internally injected witness after the PWFA stage. (d) Witness with the highest driver-to-witness energy transfer efficiency. (e) Witness with highest peak electron energy. Gray boxes highlight the spectral feature recognized as the witness. See~\cite{Note1} for detailed beam parameters.}
  \label{fig:experiment1}
\end{figure}

\emph{Experimental results.}--Figure~\ref{fig:experiment1}(a) shows the measured energy spectrum of an exemplary electron bunch generated by the LWFA stage. When the PWFA target is activated without the hydrodynamic shock (i.e., without the injection mechanism), the drive beam is collectively decelerated in a self-driven plasma wave, leaving behind a low-charge continuous energy spectrum (see Fig.~\ref{fig:experiment1}(b)). Upon insertion of the wire into the supersonic gas flow, thereby introducing a shock-induced density transition, witness electrons are injected and subsequently accelerated. The three representative witness bunches in Figure~\ref{fig:experiment1}(c)-(e) have charges between $38-53~\si{pC}$ (integrated over full bunch), a divergence at the spectral peak energy of $\sim\SI{0.1}{mrad}$ (rms of gaussian fit), and energy spreads of 3-7\% (FWHM).

\begin{figure}[t]
  \centering
  \includegraphics*[width=\linewidth]{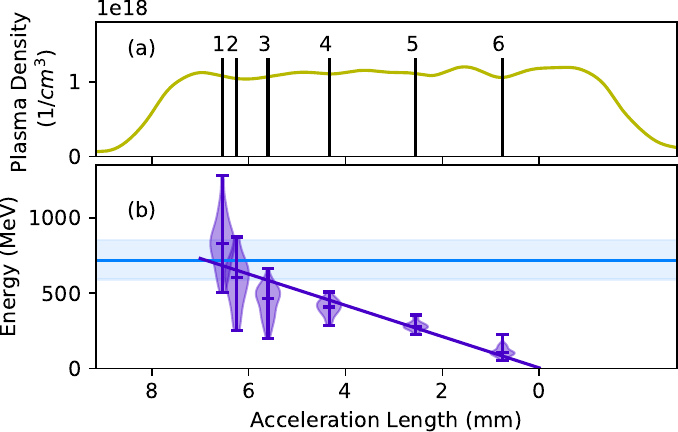}
  \caption{Tuning the witness energy by variation of the injection position. (a) Interferometric measurement of the unperturbed plasma density distribution and definition of injection positions. (b) Mean peak witness electron energies and their distribution as a function of acceleration length (violet). The zero position of the x-axis is chosen to coincide with the zero-crossing of the linear fit and corresponds to a position close to the end of the PWFA plasma target. For comparison, the drivers' mean peak electron energy and its min/max deviation are plotted in light blue. }
  \label{fig:experiment2}
\end{figure}

We control the injection point within the plasma by varying the wire's axial position, thereby adjusting the effective acceleration length. 
Figure~\ref{fig:experiment2}(a) shows the injection positions relative to the measured (unperturbed) plasma density profile. 
For the longest acceleration length, we consistently obtain witness electron energies higher than the drivers'. The highest measured electron energy gain is around $\SI{1.3}{GeV}$ or about 1.8 times the driver peak electron energy. To our knowledge, this is among the highest witness electron energy gains (relative to the driver electron energy) in all previous PWFA experiments and has only recently been surpassed~\cite{zhang2025} (see also Fig.~\ref{fig:efficiency} and the supplemental material~\footnote{See Supplemental Material at [URL will be inserted by publisher] for additional information on the experimental setup, parameters for simulations, and details on data analysis.}).  

We record the peak energy of the accelerated witness beam as a function of the remaining acceleration length. Assuming a non-evolving strength of the wakefield, we use a linear fit to these data to estimate the average accelerating gradient in the PWFA stage of $\SI{104}{GV/m}$, as shown in Figure~\ref{fig:experiment2}(b). 
For late injection positions that result in moderate witness energies, we observe increased energy stability. For example, at injection positions 4 and 5, the peak energy fluctuates by $\sim10\%$ only, consistent with the findings of our previous publication~\cite{foerster2022}. 

To interpret the maximum energy gain of a witness bunch in a beam-driven plasma accelerator, we first consider the two main limiting factors: 
(1) The accelerating gradient on the witness beam, and (2) the depletion of the driver, i.e. the point at which (a part of) the drive beam has lost all its kinetic energy to the plasma.

The transformer ratio of the wakefield relates the driver's deceleration to the witness's acceleration. 
Different definitions of the transformer ratio are used in the literature. The voltage transformer ratio is defined as the absolute value of the ratio of the peak accelerating field experienced by the witness beam to the peak decelerating field acting on the driver~\cite{bane1985}. 
While the voltage transformer ratio is directly accessible in particle-in-cell (PIC) simulations, it is not easily measurable in experiments. As an experimentally accessible quantity, one can define an energy transformer ratio as the ratio of the final peak electron energy of the witness bunch to the maximum electron energy loss of the drive bunch. This definition effectively captures the average voltage transformer ratio over the entire acceleration length. 
Note that under our experimental conditions, we can replace the electron energy loss of the driver with its peak electron energy before PWFA, because we completely decelerate (parts of) the drive bunch. Furthermore, since we inject our witness internally, and it therefore starts at zero energy, the energy transformer ratio is identical to the relative electron energy gain of the witness introduced above.

The PIC simulation in Fig~\ref{fig:setup}(a)-(c) visualizes the evolution of the longitudinal phase space of the driver and witness through the PWFA stage and illustrates factors that limit the achievable energy transformer ratio. Any energy loss of the driver prior to witness injection, such as that occurring in a long density upramp of the plasma target, reduces the energy transformer ratio. 
Also, a non-optimal witness phase position reduces the energy transformer ratio. It can arise from plasma density variations, or from the onset of driver depletion~\cite{schobel2022,hue2023a}, causing parts of the drive beam to fall back in phase and distort the wakefield structure.

To boost the energy transformer ratios beyond previously reported values, we incorporated these considerations into the design of the plasma target. We choose the plasma length to terminate the acceleration close to the point of beam breakup due to the depletion of parts of the drive bunch, as seen in Fig.~\ref{fig:setup}(c). 
Furthermore, as can be seen comparing Figure~\ref{fig:setup}(a) and (b), the driver loses only a small part of its energy in the short density up-ramp. At the same time, this section is long enough for the drive bunch to self-focus after the vacuum drift section and to drive a strong wakefield at the point of injection.

Beyond a high energy transformer ratio, our PWFA stage is characterized by an unprecedentedly high energy transfer efficiency. 
As for the transformer ratio, different definitions of energy transfer efficiency are used in the literature, describing different aspects of energy transfer between driver, plasma, and witness bunch~\cite{Lindstrom:2021eo,zhang2024,pena2024}. 
A first definition divides the integrated energy gain of the witness bunch $\Delta \varepsilon_{\text{W}}$ by the integrated energy of the incident drive bunch $\varepsilon_{\text{D}}$ and corresponds to the driver-to-witness energy transfer efficiency: 
$\eta_{D\rightarrow W} = \frac{\Delta \varepsilon_{\text{W}}}{\varepsilon_{\text{D}}}$.

In a second definition, $\Delta \varepsilon_{\text{W}}$ is divided by the integrated energy loss of the driver $\Delta \varepsilon_{\text{D}}$. Assuming that all energy lost by the driver ended up in the plasma, this can be identified as the plasma-to-witness energy transfer efficiency: 
$\eta_{P\rightarrow W} = \frac{\Delta \varepsilon_{\text{W}}}{\Delta \varepsilon_{\text{D}}}$, 
where the integrated beam energy of driver and witness is $\varepsilon_{\text{D/W}} = \int \rho_Q\left(E\right) E dE$, and $\rho_Q$ is the spectral charge density of driver and witness, respectively. $\Delta \varepsilon_{\text{D/W}}$ is the difference in integrated energy before and after the PWFA stage. 

While both metrics for efficiency capture complementary aspects of energy coupling, they may differ significantly in magnitude.  
Caution is therefore required when comparing efficiencies across different experiments or publications.

\begin{figure}[t]
  \centering
  \includegraphics*[width=\linewidth]{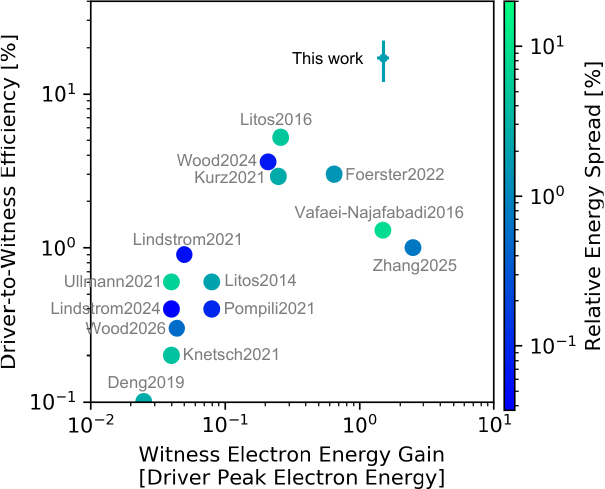}
  \caption{Comparison of published PWFA results in terms of electron energy gain of the witness normalized to the peak electron energy of the respective driver, driver-to-witness energy transfer efficiency, and relative energy spread of the witness. The error bars for this work correspond to the uncertainties discussed in the supplemental material~\cite{Note1}. Note the logarithmic scale of all three categories.}
  \label{fig:efficiency}
\end{figure}

We analyze the energy transfer efficiencies for the high-quality witness bunches with a final electron energy exceeding the driver electron energy, as clear spectral separation is ensured. The shot with the highest measured energy transfer efficiency is shown in Figure~\ref{fig:experiment1}(d). 
Its integrated energy is $\Delta \varepsilon_{\text{witness}}=66~\si{mJ}$, and that of the drive bunch is $\varepsilon_{\text{driver}}=(274~\pm~61)~\si{mJ}$ (run average and standard deviation). As we can not measure charge and energy of drive bunches before and after the PWFA stage individually, the exact integrated driver energy on the selected shot is unknown. Instead, we calculate a lower limit for the peak driver-to-witness energy transfer efficiency by assuming the integrated driver energy equals the highest value measured among all analyzed LWFA shots. This yields an efficiency:
\begin{equation*}
\eta_{D\rightarrow W}\geq\frac{\SI{66}{mJ}}{\SI{384}{mJ}}=17\%.
\end{equation*}
Using the average integrated driver energy instead, this value increases to 24\%. Averaging over all shots with witness peak energies larger than $\SI{700}{MeV}$, we find $\eta_{D\rightarrow W}=(7\pm 5)\%$ (mean and standard deviation). For a discussion of systematic uncertainties and the large shot-to-shot variations of these numbers, see~\cite{Note1}.

A value of $\eta_{D\rightarrow W} \geq 17\%$ is the highest driver-to-witness energy transfer efficiency observed in PWFA to date. Comparing the efficiencies $\eta_{D\rightarrow W}$ 
visualized in Figure~\ref{fig:efficiency}, the results of this work outperform previous experiments with either external~\cite{Litos:2016, Litos:2015ke, Lindstrom:2021eo, Kurz:2021cg, pompili2021,lindstrom2024} or internal~\cite{vafaei-najafabadi2016, deng2019,Knetsch:2021kj,Ullmann:2021ky,foerster2022} injection, by far~\cite{Note1}. 

Regarding plasma-to-witness energy transfer, we calculate a lower bound using the measured average energy remaining in the driver bunch after PWFA ($\sim \SI{100}{mJ}$) and by assuming that the spent driver energy actually contributed to driving the wakefield. 
At our experimental working point, we find
\begin{equation*}
\eta_{P\rightarrow W} \geq \frac{\SI{66}{mJ}}{(274-104)\si{mJ}} = 39\%.
\end{equation*}

Similar values of $\eta_{P\rightarrow W}$ have only been reported from experiments operated far from driver depletion~\cite{Litos:2015ke, Lindstrom:2021eo,lindstrom2024}, where the peak electron energy of the witness changes by a small fraction of the driver's peak electron energy only. Thus, while achieving very high plasma-to-witness energy transfer efficiency at their respective working points, a significantly lower fraction of the integrated drive beam energy is extracted in these previous experiments, resulting in a substantially lower driver-to-witness energy transfer efficiency.

\emph{Conclusion and Outlook.}--In summary, we have demonstrated a hybrid LWFA-PWFA operated close to driver depletion and with internal witness injection. This experiment achieves high beam quality, a high energy transformer ratio close to two, record witness energies for this scheme of $\SI{1.3}{GeV}$, and unprecedented driver-to-witness energy transfer efficiency of $\sim\SI{20}{\%}$, by far exceeding those of any previous PWFA experiment. 

These experimental results on energy transfer efficiency and energy transformer ratio represent a long-sought breakthrough. 
While a high degree of driver depletion~\cite{pena2024,zhang2024} and a high plasma-to-witness energy transfer efficiency~\cite{Lindstrom:2021eo} have already been demonstrated individually, this is the first experiment to show both simultaneously, thus enabling the high driver-to-witness energy transfer efficiency. The measured energy transformer ratio is already within the range of theoretically predicted values for the voltage transformer ratio using symmetric bunches~\cite{bane1985}. Further improvement can be expected by shaping the drive current~\cite{chen1986,lotov2005,su2023}. In experiments to date, however, very high voltage transformer ratios have been shown only in regimes with low witness electron energy gain~\cite{loisch2018,roussel2020}. 

In Figure~\ref{fig:efficiency}, we compare published PWFA results in terms of electron energy gain of the witness normalized to the peak electron energy of the respective driver, and their driver-to-witness energy transfer efficiency. This comparison reveals a correlation between the two properties, led and confirmed by our experimental results. 

Notably, our hybrid LWFA-PWFA scheme is not only at the forefront of driver-to-witness energy transfer efficiency but also generates dense, low-divergence ($\SI{\sim0.1}{mrad}$), and narrow-energy-spread (few \%, compare color bar in Fig.~\ref{fig:efficiency}) electron bunches. In terms of their low divergence and high charge density, our results are on par with, or even better than, those in a similar energy regime achieved with LWFA alone~\cite{Ke:2021hh}, enabling the practical application of plasma-based accelerators. 
Recently, several groups have shown laser-like emission and the onset of exponential gain in free-electron lasers (FEL) driven by LWFA-generated beams~\cite{labat2022,Wang:2021ko,barber2025}. Hybrid LWFA-PWFA could be considered as a compact and cost-efficient add-on for LWFA experiments, as FELs could particularly benefit from the improvements in beam parameters demonstrated here~\cite{habib2019,habib2023}. Future research on these FEL applications should address full phase space measurements, including bunch current and emittance~\cite{Heigoldt:2015cd,laberge2024,Weingartner:2012gw}.

A further application of the high-energy, high-charge-density electron bunches with low divergence could be to explore strong-field quantum electrodynamics, e.g., in the non-perturbative Breit–Wheeler pair-production regime~\cite{salgado2021}. Here, increasing electron energy is essential to achieve large quantum nonlinearity parameters, while higher charge density and lower divergence enhance event rates in these intrinsically rare processes. 
Looking further into the future, the hybrid LWFA-PWFA can also serve as a versatile miniature test bed for studying the physics of next-generation plasma-based high-energy accelerators~\cite{adli2025}, replicating all key physical processes, and being experimentally accessible and operational today.

\emph{Acknowledgments.}--We thank the Federal Republic of Germany and the Free State of Bavaria for funding the CALA infrastructure (15171 E 0002) and its operation. This work was supported by BMBF projects FIMO, MACLIP, and HyBright. 
F.M.F.\ is part of the Max Planck School of Photonics, supported by BMBF, Max Planck Society, and Fraunhofer Society. B.H., T.H., M.C., and F.A.H. are supported by the European Research Council (ERC) under the EU Horizon 2020 research and innovation programme (NeXource: Next-generation Plasma-based Electron Beam Sources for High-brightness Photon Science, ERC Grant agreement No. 865877.


\bibliography{references}

@article{Blumenfeld:2007ja,
author = {Blumenfeld, I. and Clayton, C. E. and Decker, F. J. and Hogan, M. J. and Huang, C. K. and Ischebeck, R. and Iverson, R. and Joshi, C. and Katsouleas, T. and Kirby, N. and Lu, W. and Marsh, K. A. and Mori, W. B. and Muggli, P. and Oz, E. and Siemann, R. H. and Walz, D. and Zhou, M. M.},
title = {{Energy doubling of 42 GeV electrons in a metre-scale plasma wakefield accelerator}},
journal = {Nature},
year = {2007},
volume = {445},
number = {7129},
pages = {741--744},
month = feb
}

@article{Maier:2020kh,
author = {Maier, Andreas R and Delbos, Niels M and Eichner, Timo and H{\"u}bner, Lars and Jalas, S{\"o}ren and Jeppe, Laurids and Jolly, Spencer W and Kirchen, Manuel and Leroux, Vincent and Messner, Philipp and Schnepp, Matthias and Trunk, Maximilian and Walker, Paul A and Werle, Christian and Winkler, Paul},
title = {{Decoding Sources of Energy Variability in a Laser-Plasma Accelerator}},
journal = {Physical Review X},
year = {2020},
volume = {10},
number = {3},
pages = {031039},
month = aug
}

@article{Lindstrom:2021eo,
  title = {Energy-Spread Preservation and High Efficiency in a Plasma-Wakefield Accelerator},
  author = {Lindstr\o{}m, C. A. and Garland, J. M. and Schr\"oder, S. and Boulton, L. and Boyle, G. and Chappell, J. and D'Arcy, R. and Gonzalez, P. and Knetsch, A. and Libov, V. and Loisch, G. and Martinez de la Ossa, A. and Niknejadi, P. and P\~oder, K. and Schaper, L. and Schmidt, B. and Sheeran, B. and Wesch, S. and Wood, J. and Osterhoff, J.},
  journal = {Phys. Rev. Lett.},
  volume = {126},
  issue = {1},
  pages = {014801},
  numpages = {6},
  year = {2021},
  month = Jan
  }

@article{Kurz:2021cg,
author = {Kurz, T and Heinemann, T and Gilljohann, M F and Chang, Y Y and Cabada{\u g}, J P Couperus and Debus, A. and Kononenko, O and Pausch, R and Sch{\"o}bel, S and Assmann, R W and Bussmann, M. and Ding, H and G{\"o}tzfried, J and K{\"o}hler, A and Raj, G. and Schindler, S and Steiniger, K and Zarini, O and Corde, S. and D{\"o}pp, A and Hidding, B. and Karsch, S. and Schramm, U. and de la Ossa, A Martinez and Irman, A.},
title = {{Demonstration of a compact plasma accelerator powered by laser-accelerated electron beams}},
journal = {Nature Communications},
volume = {12},
year = {2021},
pages = {2895},
month = may
}

@article{CouperusCabadag:2021gj,
  title = {Gas-dynamic density downramp injection in a beam-driven plasma wakefield accelerator},
  author = {Couperus Cabada\ifmmode \breve{g}\else \u{g}\fi{}, J. P. and Pausch, R. and Sch\"obel, S. and Bussmann, M. and Chang, Y.-Y. and Corde, S. and Debus, A. and Ding, H. and D\"opp, A. and Foerster, F. M. and Gilljohann, M. and Haberstroh, F. and Heinemann, T. and Hidding, B. and Karsch, S. and Koehler, A. and Kononenko, O. and Knetsch, A. and Kurz, T. and Martinez de la Ossa, A. and Nutter, A. and Raj, G. and Steiniger, K. and Schramm, U. and Ufer, P. and Irman, A.},
  journal = {Phys. Rev. Research},
  volume = {3},
  issue = {4},
  pages = {L042005},
  numpages = {9},
  year = {2021},
  month = Oct
}

@article{Wang:2021ko,
author = {Wang, Wentao and Feng, Ke and Ke, Lintong and Yu, Changhai and Xu, Yi and Qi, Rong and Chen, Yu and Qin, Zhiyong and Zhang, Zhijun and Fang, Ming and Liu, Jiaqi and Jiang, Kangnan and Wang, Hao and Wang, Cheng and Yang, Xiaojun and Wu, Fenxiang and Leng, Yuxin and Liu, Jiansheng and Li, Ruxin and Xu, Zhizhan},
title = {{Free-electron lasing at 27 nanometres based on a laser wakefield accelerator}},
journal = {Nature},
year = {2021},
volume = {595},
issue = {7868},
pages = {516-520},
month = jul
}

@article{Knetsch:2021kj,
  title = {Stable witness-beam formation in a beam-driven plasma cathode},
  author = {Knetsch, A. and Sheeran, B. and Boulton, L. and Niknejadi, P. and P\~oder, K. and Schaper, L. and Zeng, M. and Bohlen, S. and Boyle, G. and Br\"ummer, T. and Chappell, J. and D'Arcy, R. and Diederichs, S. and Foster, B. and Garland, M. J. and Gonzalez Caminal, P. and Hidding, B. and Libov, V. and Lindstr\o{}m, C. A. and Martinez de la Ossa, A. and Meisel, M. and Parikh, T. and Schmidt, B. and Schr\"oder, S. and Tauscher, G. and Wesch, S. and Winkler, P. and Wood, J. C. and Osterhoff, J.},
  journal = {Phys. Rev. Accel. Beams},
  volume = {24},
  issue = {10},
  pages = {101302},
  numpages = {10},
  year = {2021},
  month = Oct
}

@article{Ullmann:2021ky,
author = {Ullmann, D and Scherkl, P and Knetsch, A and Heinemann, T and Sutherland, A and Habib, A F and Karger, O S and Beaton, A and Manahan, G. G. and Deng, A and Andonian, G and Litos, M D and OShea, B D and Cary, J. R. and Hogan, M. J. and Yakimenko, V. and Rosenzweig, J. B. and Hidding, B.},
title = {{All-optical density downramp injection in electron-driven plasma wakefield accelerators}},
journal = {Phys. Rev. Research},
year = {2021},
volume = {3},
pages = {043163},
month = dec
}

@article{Gonsalves:2019ht,
author = {Gonsalves, A. J. and Nakamura, K. and Daniels, J and Benedetti, C and Pieronek, C and de Raadt, T C H and Steinke, S and Bin, J H and Bulanov, S. S. and van Tilborg, J. and Geddes, C. G. R. and Schroeder, C. B. and Toth, Cs and Esarey, E. and Swanson, K and Fan-Chiang, L and Bagdasarov, G and Bobrova, N and Gasilov, V and Korn, G and Sasorov, P and Leemans, W. P.},
title = {{Petawatt Laser Guiding and Electron Beam Acceleration to 8 GeV in a Laser-Heated Capillary Discharge Waveguide}},
journal = {Physical Review Letters},
year = {2019},
volume = {122},
number = {8},
pages = {084801},
month = feb
}

@article{Gotzfried:2020da,
author = {G{\"o}tzfried, J and D{\"o}pp, A and Gilljohann, M F and Foerster, F M and Ding, H and Schindler, S and Schilling, G and Buck, A. and Veisz, L. and Karsch, S.},
title = {{Physics of High-Charge Electron Beams in Laser-Plasma Wakefields}},
journal = {Physical Review X},
year = {2020},
volume = {10},
number = {4},
pages = {041015},
month = oct
}

@article{Heigoldt:2015cd,
author = {Heigoldt, M and Popp, A. and Khrennikov, K and Wenz, J and Chou, S W and Karsch, S. and Bajlekov, S. I. and Hooker, S. M. and Schmidt, B},
title = {{Temporal evolution of longitudinal bunch profile in a laser wakefield accelerator}},
journal = {Physical Review Special Topics-Accelerators and Beams},
year = {2015},
volume = {18},
number = {12},
pages = {121302},
month = dec
}

@article{Couperus:2017bg,
author = {Couperus, J P and Pausch, R and K{\"o}hler, A and Zarini, O and Kr{\"a}mer, J M and Garten, M and Huebl, A and Gebhardt, R and Helbig, U and Bock, S and Zeil, K. and Debus, A. and Bussmann, M. and Schramm, U. and Irman, A.},
title = {{Demonstration of a beam loaded nanocoulomb-class laser wakefield accelerator}},
journal = {Nature Communications},
year = {2017},
volume = {8},
issue = {1},
number = {487},
pages = {1--7},
month = aug
}

@article{Chou:2016ei,
author = {Chou, S and Xu, J and Khrennikov, K and Cardenas, D E and Wenz, J and Heigoldt, M and Hofmann, L and Veisz, L. and Karsch, S.},
title = {{Collective Deceleration of Laser-Driven Electron Bunches}},
journal = {Physical Review Letters},
year = {2016},
volume = {117},
number = {14},
pages = {144801},
month = sep
}

@article{Weingartner:2012gw,
author = {Weingartner, R. and Raith, S and Popp, A. and Chou, S and Wenz, J and Khrennikov, K and Heigoldt, M and Maier, A and Kajumba, N and Fuchs, M. and Zeitler, B. and Krausz, F. and Karsch, S. and Gruner, F.},
title = {{Ultralow emittance electron beams from a laser-wakefield accelerator}},
journal = {Physical Review Special Topics-Accelerators and Beams},
year = {2012},
volume = {15},
number = {11},
pages = {111302},
month = nov
}

@article{Hidding:2012ep,
author = {Hidding, B. and Pretzler, G. and Rosenzweig, J. B. and K{\"o}nigstein, T and Schiller, D and Bruhwiler, D L},
title = {{Ultracold Electron Bunch Generation via Plasma Photocathode Emission and Acceleration in a Beam-Driven Plasma Blowout}},
journal = {Physical Review Letters},
year = {2012},
volume = {108},
number = {3},
pages = {035001},
month = jan
}

@article{MartinezdelaOssa:2015es,
author = {Martinez de la Ossa, A and Mehrling, T J and Schaper, L and Streeter, M. J. V. and Osterhoff, J.},
title = {{Wakefield-induced ionization injection in beam-driven plasma accelerators}},
journal = {Physics of Plasmas},
year = {2015},
volume = {22},
number = {9},
pages = {093107},
month = sep
}

@article{Gilljohann:2019kc,
  title = {Direct Observation of Plasma Waves and Dynamics Induced by Laser-Accelerated Electron Beams},
  author = {Gilljohann, M. F. and Ding, H. and D\"opp, A. and G\"otzfried, J. and Schindler, S. and Schilling, G. and Corde, S. and Debus, A. and Heinemann, T. and Hidding, B. and Hooker, S. M. and Irman, A. and Kononenko, O. and Kurz, T. and Martinez de la Ossa, A. and Schramm, U. and Karsch, S.},
  journal = {Phys. Rev. X},
  volume = {9},
  issue = {1},
  pages = {011046},
  numpages = {13},
  year = {2019},
  month = Mar
}

@article{Lehe:2016dn,
author = {Lehe, Remi and Kirchen, Manuel and Andriyash, Igor A and Godfrey, Brendan B and Vay, Jean-Luc},
title = {{A spectral, quasi-cylindrical and dispersion-free Particle-In-Cell algorithm}},
journal = {Computer Physics Communications},
year = {2016},
volume = {203},
pages = {66--82}
}

@article{Ke:2021hh,
author = {Ke, L T and Feng, K and Wang, W. T. and Qin, Z Y and Yu, C H and Wu, Y and Chen, Y and Qi, R and Zhang, Z J and Xu, Y and Yang, X J and Leng, Y. X. and Liu, J. S. and Li, R. X. and Xu, Z. Z.},
title = {{Near-GeV Electron Beams at a Few Per-Mille Level from a Laser Wakefield Accelerator via Density-Tailored Plasma}},
journal = {Physical Review Letters},
year = {2021},
volume = {126},
pages = {214801},
month = may
}

@article{Litos:2015ke,
author = {Litos, M and Adli, E and An, W and Clarke, C I and Clayton, C. E. and Corde, S. and Delahaye, J P and England, R J and Fisher, A S and Frederico, J and Gessner, S and Green, S Z and Hogan, M. J. and Joshi, C. and Lu, W. and Marsh, K. A. and Mori, W. B. and Muggli, P. and Vafaei-Najafabadi, N and Walz, D. and White, G and Wu, Z and Yakimenko, V. and Yocky, G.},
title = {{High-efficiency acceleration of an electron beam in a plasma wakefield accelerator}},
journal = {Nature},
year = {2014},
volume = {515},
number = {7525},
pages = {92--95},
month = apr
}

@article{Litos:2016,
	title = {9 {GeV} energy gain in a beam-driven plasma wakefield accelerator},
	author = {M Litos and E Adli and J M Allen and W An and C I Clarke and S Corde and C E Clayton and J Frederico and S J Gessner and S Z Green and M J Hogan and C Joshi and W Lu and K A Marsh and W B Mori and M Schmeltz and N Vafaei-Najafabadi and V Yakimenko},
	year = 2016,
	month = feb,
	volume = {58},
	number = {3},
	pages = {034017},
	journal = {Plasma Physics and Controlled Fusion},
}

@article{MartinezdelaOssa:2019bm,
author = {Martinez de la Ossa, A and Assmann, R W and Bussmann, M. and Corde, S. and Couperus Cabada{\u g}, J P and Debus, A. and D{\"o}pp, A and Ferran Pousa, A and Gilljohann, M F and Heinemann, T and Hidding, B. and Irman, A. and Karsch, S. and Kononenko, O and Kurz, T and Osterhoff, J. and Pausch, R and Sch{\"o}bel, S and Schramm, U.},
title = {{Hybrid LWFA|PWFA staging as a beam energy and brightness transformer: conceptual design and simulations}},
journal = {Philosophical Transactions of the Royal Society A: Mathematical, Physical and Engineering Sciences},
year = {2019},
volume = {377},
number = {2151},
pages = {20180175}
}

@article{Adli:2018jza,
author = {Adli, E and Ahuja, A and Apsimon, O and Apsimon, R and Bachmann, A M and Barrientos, D and Batsch, F and Bauche, J and Olsen, V K Berglyd and Bernardini, M and Bohl, T and Bracco, C and Braunm{\"u}ller, F and Burt, G and Buttensch{\"o}n, B and Caldwell, A. and Cascella, M and Chappell, J and Chevallay, E and Chung, M and Cooke, D and Damerau, H and Deacon, L and Deubner, L H and Dexter, A and Doebert, S and Farmer, J and Fedosseev, V N and Fiorito, R and Fonseca, R. A. and Friebel, F and Garolfi, L and Gessner, S and Gorgisyan, I and Gorn, A A and Granados, E and Grulke, O and Gschwendtner, E and Hansen, J and Helm, A and Henderson, J R and H{\"u}ther, M and Ibison, M and Jensen, L and Jolly, S and Keeble, F and Kim, S Y and Kraus, F and Li, Y and Liu, S and Lopes, N. and Lotov, K V and Brun, L Maricalva and Martyanov, M and Mazzoni, S and Godoy, D Medina and Minakov, V A and Mitchell, J and Molendijk, J C and Moody, J T and Moreira, M and Muggli, P. and Oz, E. and Pasquino, C and Pardons, A and Asmus, F Pe{\~n}a and Pepitone, K and Perera, A and Petrenko, A and Pitman, S and Pukhov, A. and Rey, S and Rieger, K and Ruhl, H. and Schmidt, J S and Shalimova, I A and Sherwood, P and Silva, L. O. and Soby, L and Sosedkin, A P and Speroni, R and Spitsyn, R I and Tuev, P V and Turner, M and Velotti, F and Verra, L and Verzilov, V A and Vieira, J. and Welsch, C P and Williamson, B and Wing, M and Woolley, B and Xia, G},
collaboration = {AWAKE},
title = {{Acceleration of electrons in the plasma wakefield of a proton bunch}},
journal = {Nature},
year = {2018},
volume = {561},
issue = {7723},
pages = {363--367},
month = sep
}

@article{pompili2021,
title = {Energy spread minimization in a beam-driven plasma wakefield accelerator},
issn = {1745-2481},
url = {https://www.nature.com/articles/s41567-020-01116-9},
doi = {10.1038/s41567-020-01116-9},
urldate = {2021-01-27},
journal = {Nature Physics},
author = {Pompili, R. and Alesini, D. and Anania, M. P. and Behtouei, M. and Bellaveglia, M. and Biagioni, A. and Bisesto, F. G. and Cesarini, M. and Chiadroni, E. and Cianchi, A. and Costa, G. and Croia, M. and Del Dotto, A. and Di Giovenale, D. and Diomede, M. and Dipace, F. and Ferrario, M. and Giribono, A. and Lollo, V. and Magnisi, L. and Marongiu, M. and Mostacci, A. and Piersanti, L. and Di Pirro, G. and Romeo, S. and Rossi, A. R. and Scifo, J. and Shpakov, V. and Vaccarezza, C. and Villa, F. and Zigler, A.},
month = jan,
year = {2021},
pages = {499–503},
volume = {17}
}

@article{foerster2022,
  title = {Stable and {{High-Quality Electron Beams}} from {{Staged Laser}} and {{Plasma Wakefield Accelerators}}},
  author = {Foerster, F. M. and D{\"o}pp, A. and Haberstroh, F. and v. Grafenstein, K. and Campbell, D. and Chang, Y.-Y. and Corde, S. and Couperus Cabada{\u g}, J. P. and Debus, A. and Gilljohann, M. F. and Habib, A. F. and Heinemann, T. and Hidding, B. and Irman, A. and Irshad, F. and Knetsch, A. and Kononenko, O. and {Martinez de la Ossa}, A. and Nutter, A. and Pausch, R. and Schilling, G. and Schletter, A. and Sch{\"o}bel, S. and Schramm, U. and Travac, E. and Ufer, P. and Karsch, S.},
  year = {2022},
  month = nov,
  journal = {Physical Review X},
  volume = {12},
  number = {4},
  pages = {041016},
  publisher = {{American Physical Society}},
  doi = {10.1103/PhysRevX.12.041016}
}

@article{schobel2022,
  title = {Effect of Driver Charge on Wakefield Characteristics in a Plasma Accelerator Probed by Femtosecond Shadowgraphy},
  author = {Sch{\"o}bel, Susanne and Pausch, Richard and Chang, Yen-Yu and Corde, S{\'e}bastien and Couperus Cabada{\u g}, Jurjen and Debus, Alexander and Ding, Hao and D{\"o}pp, Andreas and Foerster, F Moritz and Gilljohann, Max and Haberstroh, Florian and Heinemann, Thomas and Hidding, Bernhard and Karsch, Stefan and K{\"o}hler, Alexander and Kononenko, Olena and Kurz, Thomas and Nutter, Alastair and Steiniger, Klaus and Ufer, Patrick and {Martinez de la Ossa}, Alberto and Schramm, Ulrich and Irman, Arie},
  year = {2022},
  month = aug,
  journal = {New Journal of Physics},
  volume = {24},
  number = {8},
  pages = {083034},
  issn = {1367-2630},
  doi = {10.1088/1367-2630/ac87c9}
}

@article{zarini2022,
  title = {Multioctave High-Dynamic Range Optical Spectrometer for Single-Pulse, Longitudinal Characterization of Ultrashort Electron Bunches},
  author = {Zarini, Omid and Cabada{\u g}, Jurjen Couperus and Chang, Yen-Yu and K{\"o}hler, Alexander and Kurz, Thomas and Sch{\"o}bel, Susanne and Seidel, Wolfgang and Bussmann, Michael and Schramm, Ulrich and Irman, Arie and Debus, Alexander},
  year = {2022},
  month = jan,
  journal = {Physical Review Accelerators and Beams},
  volume = {25},
  number = {1},
  pages = {012801},
  doi = {10.1103/PhysRevAccelBeams.25.012801}
}

@article{labat2022,
  title = {Seeded Free-Electron Laser Driven by a Compact Laser Plasma Accelerator},
  author = {Labat, Marie and Cabada{\u g}, Jurjen Couperus and Ghaith, Amin and Irman, Arie and Berlioux, Anthony and Berteaud, Philippe and Blache, Fr{\'e}d{\'e}ric and Bock, Stefan and Bouvet, Fran{\c c}ois and Briquez, Fabien and Chang, Yen-Yu and Corde, S{\'e}bastien and Debus, Alexander and De Oliveira, Carlos and Duval, Jean-Pierre and Dietrich, Yannick and El Ajjouri, Moussa and Eisenmann, Christoph and Gautier, Julien and Gebhardt, Ren{\'e} and Grams, Simon and Helbig, Uwe and Herbeaux, Christian and Hubert, Nicolas and Kitegi, Charles and Kononenko, Olena and Kuntzsch, Michael and LaBerge, Maxwell and L{\^e}, St{\'e}phane and Leluan, Bruno and Loulergue, Alexandre and Malka, Victor and Marteau, Fabrice and Guyen, Manh Huy N. and {Oumbarek-Espinos}, Driss and Pausch, Richard and Pereira, Damien and P{\"u}schel, Thomas and Ricaud, Jean-Paul and Rommeluere, Patrick and Roussel, El{\'e}onore and Rousseau, Pascal and Sch{\"o}bel, Susanne and Sebdaoui, Mourad and Steiniger, Klaus and Tavakoli, Keihan and Thaury, C{\'e}dric and Ufer, Patrick and Vall{\'e}au, Mathieu and Vandenberghe, Marc and V{\'e}t{\'e}ran, Jos{\'e} and Schramm, Ulrich and Couprie, Marie-Emmanuelle},
  year = {2022},
  month = dec,
  journal = {Nature Photonics},
  pages = {150–156},
  volume = {17}, 
  issn = {1749-4893},
  doi = {10.1038/s41566-022-01104-w}
}

@article{hidding2010,
  title = {Monoenergetic {{Energy Doubling}} in a {{Hybrid Laser-Plasma Wakefield Accelerator}}},
  author = {Hidding, B. and K{\"o}nigstein, T. and Osterholz, J. and Karsch, S. and Willi, O. and Pretzler, G.},
  year = {2010},
  month = may,
  journal = {Physical Review Letters},
  volume = {104},
  number = {19},
  pages = {195002},
  publisher = {{American Physical Society}},
  doi = {10.1103/PhysRevLett.104.195002}
}

@article{hidding2023,
  title = {Progress in {{Hybrid Plasma Wakefield Acceleration}}},
  author = {Hidding, Bernhard and Assmann, Ralph and Bussmann, Michael and Campbell, David and Chang, Yen-Yu and Corde, S{\'e}bastien and Cabada{\u g}, Jurjen Couperus and Debus, Alexander and D{\"o}pp, Andreas and Gilljohann, Max and G{\"o}tzfried, J. and Foerster, F. Moritz and Haberstroh, F. Florian and Habib, Fahim and Heinemann, Thomas and Hollatz, Dominik and Irman, Arie and Kaluza, Malte and Karsch, Stefan and Kononenko, Olena and Knetsch, Alexander and Kurz, Thomas and Kuschel, Stephan and K{\"o}hler, Alexander and de la Ossa, Alberto Martinez and Nutter, Alastair and Pausch, Richard and Raj, Gaurav and Schramm, Ulrich and Sch{\"o}bel, Susanne and Seidel, Andreas and Steiniger, Klaus and Ufer, Patrick and Yeung, Mark and Zarini, Omid and Zepf, Matt},
  year = {2023},
  month = jan,
  journal = {Photonics},
  volume = {10},
  number = {2},
  pages = {99},
  issn = {2304-6732},
  doi = {10.3390/photonics10020099}
}

@article{esarey2009,
  title = {Physics of Laser-Driven Plasma-Based Electron Accelerators},
  author = {Esarey, E. and Schroeder, C. B. and Leemans, W. P.},
  year = {2009},
  month = aug,
  journal = {Reviews of Modern Physics},
  volume = {81},
  number = {3},
  pages = {1229--1285},
  issn = {0034-6861, 1539-0756},
  doi = {10.1103/RevModPhys.81.1229},
}

@article{tajima1979,
  title = {Laser {{Electron Accelerator}}},
  author = {Tajima, T. and Dawson, J. M.},
  year = {1979},
  month = jul,
  journal = {Physical Review Letters},
  volume = {43},
  number = {4},
  pages = {267--270},
  issn = {0031-9007},
  doi = {10.1103/PhysRevLett.43.267},
}

@misc{FBPIC,
  year = {2016},
  title = {FBPIC documentation},
  howpublished = {\url{https://fbpic.github.io}},
  note = {Accessed: 2023-07-30}
}

@article{su2023,
  title = {Optimization of Transformer Ratio and Beam Loading in a Plasma Wakefield Accelerator with a Structure-Exploiting Algorithm},
  author = {Su, Q. and Larson, J. and Dalichaouch, T. N. and Li, F. and An, W. and Hildebrand, L. and Zhao, Y. and Decyk, V. and Alves, P. and Wild, S. M. and Mori, W. B.},
  year = {2023},
  month = may,
  journal = {Physics of Plasmas},
  volume = {30},
  number = {5},
  pages = {053108},
  issn = {1070-664X, 1089-7674},
  doi = {10.1063/5.0142940}
}

@article{darcy2019,
  title = {{{FLASHForward}}: Plasma Wakefield Accelerator Science for High-Average-Power Applications},
  shorttitle = {{{FLASHForward}}},
  author = {D'Arcy, R. and Aschikhin, A. and Bohlen, S. and Boyle, G. and Br{\"u}mmer, T. and Chappell, J. and Diederichs, S. and Foster, B. and Garland, M. J. and Goldberg, L. and Gonzalez, P. and Karstensen, S. and Knetsch, A. and Kuang, P. and Libov, V. and Ludwig, K. and {Martinez de la Ossa}, A. and Marutzky, F. and Meisel, M. and Mehrling, T. J. and Niknejadi, P. and P{\~o}der, K. and Pourmoussavi, P. and Quast, M. and R{\"o}ckemann, J. -H. and Schaper, L. and Schmidt, B. and Schr{\"o}der, S. and Schwinkendorf, J. -P. and Sheeran, B. and Tauscher, G. and Wesch, S. and Wing, M. and Winkler, P. and Zeng, M. and Osterhoff, J.},
  year = {2019},
  month = jun,
  journal = {Philosophical Transactions of the Royal Society A: Mathematical, Physical and Engineering Sciences},
  volume = {377},
  number = {2151},
  pages = {20180392},
  publisher = {{Royal Society}},
  doi = {10.1098/rsta.2018.0392}
}

@article{pena2024,
  title = {Energy Depletion and Re-Acceleration of Driver Electrons in a Plasma-Wakefield Accelerator},
  author = {Pe{\~n}a, F. and Lindstr{\o}m, C. A. and Beinortait{\.e}, J. and Bj{\"o}rklund Svensson, J. and Boulton, L. and Diederichs, S. and Foster, B. and Garland, J. M. and Gonz{\'a}lez Caminal, P. and Loisch, G. and Schr{\"o}der, S. and Th{\'e}venet, M. and Wesch, S. and Wood, J. C. and Osterhoff, J. and D'Arcy, R.},
  year = {2024},
  month = nov,
  journal = {Physical Review Research},
  volume = {6},
  number = {4},
  pages = {043090},
  publisher = {American Physical Society},
  doi = {10.1103/PhysRevResearch.6.043090}
}

@article{lindstrom2024,
  title = {Emittance Preservation in a Plasma-Wakefield Accelerator},
  author = {Lindstr{\o}m, C. A. and Beinortait{\.e}, J. and Bj{\"o}rklund Svensson, J. and Boulton, L. and Chappell, J. and Diederichs, S. and Foster, B. and Garland, J. M. and Gonz{\'a}lez Caminal, P. and Loisch, G. and Pe{\~n}a, F. and Schr{\"o}der, S. and Th{\'e}venet, M. and Wesch, S. and Wing, M. and Wood, J. C. and D'Arcy, R. and Osterhoff, J.},
  year = {2024},
  month = jul,
  journal = {Nature Communications},
  volume = {15},
  number = {1},
  pages = {6097},
  publisher = {Nature Publishing Group},
  issn = {2041-1723},
  doi = {10.1038/s41467-024-50320-1}
}

@article{zhang2024,
  title = {Generation of Meter-Scale Hydrogen Plasmas and Efficient, Pump-Depletion-Limited Wakefield Excitation Using 10 {{GeV}} Electron Bunches},
  author = {Zhang, C and Storey, D and San Miguel Claveria, P and Nie, Z and Marsh, K A and Hogan, M and Mori, W B and Adli, E and An, W and Ariniello, R and Cao, G J and Clarke, C and Corde, S and Dalichaouch, T and Doss, C E and Emma, C and Ekerfelt, H and Gerstmayr, E and Gessner, S and Hansel, C and Knetsch, A and Lee, V and Li, F and Litos, M and O'Shea, B and White, G and Yocky, G and Zakharova, V and Joshi, C},
  year = {2024},
  month = feb,
  journal = {Plasma Physics and Controlled Fusion},
  volume = {66},
  number = {2},
  pages = {025013},
  issn = {0741-3335, 1361-6587},
  doi = {10.1088/1361-6587/ad1ae4}
}

@article{roussel2020,
  title = {Single {{Shot Characterization}} of {{High Transformer Ratio Wakefields}} in {{Nonlinear Plasma Acceleration}}},
  author = {Roussel, R. and Andonian, G. and Lynn, W. and Sanwalka, K. and Robles, R. and Hansel, C. and Deng, A. and Lawler, G. and Rosenzweig, J. B. and Ha, G. and Seok, J. and Power, J. G. and Conde, M. and Wisniewski, E. and Doran, D. S. and Whiteford, C. E.},
  year = {2020},
  month = jan,
  journal = {Physical Review Letters},
  volume = {124},
  number = {4},
  pages = {044802},
  publisher = {American Physical Society},
  doi = {10.1103/PhysRevLett.124.044802}
}

@article{deng2019,
  title = {Generation and Acceleration of Electron Bunches from a Plasma Photocathode},
  author = {Deng, A. and Karger, O. S. and Heinemann, T. and Knetsch, A. and Scherkl, P. and Manahan, G. G. and Beaton, A. and Ullmann, D. and Wittig, G. and Habib, A. F. and Xi, Y. and Litos, M. D. and O'Shea, B. D. and Gessner, S. and Clarke, C. I. and Green, S. Z. and Lindstr{\o}m, C. A. and Adli, E. and Zgadzaj, R. and Downer, M. C. and Andonian, G. and Murokh, A. and Bruhwiler, D. L. and Cary, J. R. and Hogan, M. J. and Yakimenko, V. and Rosenzweig, J. B. and Hidding, B.},
  year = {2019},
  month = nov,
  journal = {Nature Physics},
  volume = {15},
  number = {11},
  pages = {1156--1160},
  publisher = {Nature Publishing Group},
  issn = {1745-2481},
  doi = {10.1038/s41567-019-0610-9}
}

@article{loisch2018,
  title = {Observation of {{High Transformer Ratio Plasma Wakefield Acceleration}}},
  author = {Loisch, Gregor and Asova, Galina and Boonpornprasert, Prach and Brinkmann, Reinhard and Chen, Ye and Engel, Johannes and Good, James and Gross, Matthias and Gr{\"u}ner, Florian and Huck, Holger and Kalantaryan, Davit and Krasilnikov, Mikhail and Lishilin, Osip and {de la Ossa}, Alberto Martinez and Mehrling, Timon J. and Melkumyan, David and Oppelt, Anne and Osterhoff, Jens and Qian, Houjun and Renier, Yves and Stephan, Frank and Tenholt, Carmen and Wohlfarth, Valentin and Zhao, Quantang},
  year = {2018},
  month = aug,
  journal = {Physical Review Letters},
  volume = {121},
  number = {6},
  pages = {064801},
  issn = {0031-9007, 1079-7114},
  doi = {10.1103/PhysRevLett.121.064801}
}

@article{vafaei-najafabadi2016,
  title = {Evidence for High-Energy and Low-Emittance Electron Beams Using Ionization Injection of Charge in a Plasma Wakefield Accelerator},
  author = {{Vafaei-Najafabadi}, N and An, W and Clayton, C E and Joshi, C and Marsh, K A and Mori, W B and Welch, E C and Lu, W and Adli, E and Allen, J and Clarke, C I and Corde, S and Frederico, J and Gessner, S J and Green, S Z and Hogan, M J and Litos, M D and Yakimenko, V},
  year = {2016},
  month = mar,
  journal = {Plasma Physics and Controlled Fusion},
  volume = {58},
  number = {3},
  pages = {034009},
  issn = {0741-3335, 1361-6587},
  doi = {10.1088/0741-3335/58/3/034009}
}

@misc{adli2025,
  title = {{{HALHF}}: A Hybrid, Asymmetric, Linear {{Higgs}} Factory Using Plasma- and {{RF-based}} Acceleration},
  shorttitle = {{{HALHF}}},
  author = {Adli, Erik and Appleby, Joshua and Barklow, Timothy L. and Biagini, Marica and Svensson, Jonas Bj{\"o}rklund and Berggren, Mikael and Bettoni, Simone and Boogert, Stewart and Burrows, Philip and Caldwell, Allen and Chen, Jian Bin Ben and Cilento, Vera and Corner, Laura and D'Arcy, Richard and Doebert, Steffen and Dou, Wang and Drobniak, Pierre and Dyson, Calvin and Farrington, Sinead and Farmer, John and {Faus-Golfe}, Angeles and Formela, Manuel and Formenti, Arianne and Forrester, Louis and Foster, Brian and Gao, Jie and Gessner, Spencer and Hamann, Niclas and Harrison, Alexander and Hogan, Mark J. and H{\o}rlyk, Eir Eline and Kalvik, Daniel and Laudrain, Antoine and Lehe, Reme and Leemans, Wim and Lindstr{\o}m, Carl A. and List, Benno and List, Jenny and Lu, Xueying and Mactavish, Edward and Maslov, Vasyl and Nanni, Emilio and Osborne, John and Osterhoff, Jens and Pe{\~n}a, Felipe and Pick, Gudrid Moortgat and P{\~o}der, Kristjan and Reuter, Jurgen and Samoilenko, Dmitrii and Sedlackova, Nela and Seryi, Andrei and Sjobak, Kyrre and Sloan, Terry and Stapnes, Steinar and Garcia, Rogelio Tomas and Titov, Maxim and Trautwein, Malte and Th{\'e}venet, Maxence and Walker, Nicholas J. and Wenskat, Marc and Wing, Matthew and Wood, Jonathan},
  year = {2025},
  publisher = {arXiv},
  doi = {10.48550/ARXIV.2503.19880}
}

@article{laberge2024,
  title = {Revealing the Three-Dimensional Structure of Microbunched Plasma-Wakefield-Accelerated Electron Beams},
  author = {LaBerge, Maxwell and Bowers, Brant and Chang, Yen-Yu and Couperus Cabada{\u g}, Jurjen and Debus, Alexander and Hannasch, Andrea and Pausch, Richard and Sch{\"o}bel, Susanne and Tiebel, Jessica and Ufer, Patrick and Willmann, Anna and Zarini, Omid and Zgadzaj, Rafal and Lumpkin, Alex H. and Schramm, Ulrich and Irman, Arie and Downer, M. C.},
  year = {2024},
  month = jul,
  journal = {Nature Photonics},
  pages = {1--8},
  publisher = {Nature Publishing Group},
  issn = {1749-4893},
  doi = {10.1038/s41566-024-01475-2}
}

@article{shrock2024,
  title = {Guided {{Mode Evolution}} and {{Ionization Injection}} in {{Meter-Scale Multi-GeV Laser Wakefield Accelerators}}},
  author = {Shrock, J. E. and Rockafellow, E. and Miao, B. and Le, M. and Hollinger, R. C. and Wang, S. and Gonsalves, A. J. and Picksley, A. and Rocca, J. J. and Milchberg, H. M.},
  year = {2024},
  month = jul,
  journal = {Physical Review Letters},
  volume = {133},
  number = {4},
  pages = {045002},
  publisher = {American Physical Society},
  doi = {10.1103/PhysRevLett.133.045002}
}

@article{picksley2024,
  title = {Matched {{Guiding}} and {{Controlled Injection}} in {{Dark-Current-Free}}, 10-{{GeV-Class}}, {{Channel-Guided Laser-Plasma Accelerators}}},
  author = {Picksley, A. and Stackhouse, J. and Benedetti, C. and Nakamura, K. and Tsai, H. E. and Li, R. and Miao, B. and Shrock, J. E. and Rockafellow, E. and Milchberg, H. M. and Schroeder, C. B. and {van Tilborg}, J. and Esarey, E. and Geddes, C. G. R. and Gonsalves, A. J.},
  year = {2024},
  month = dec,
  journal = {Physical Review Letters},
  volume = {133},
  number = {25},
  pages = {255001},
  publisher = {American Physical Society},
  doi = {10.1103/PhysRevLett.133.255001}
}

@techreport{author2016,
  author       = {SLAC National Accelerator Lab.},
  title        = {Technical Design Report for the FACET-II Project at SLAC National Accelerator Laboratory},
  institution  = {SLAC National Accelerator Lab., Menlo Park, CA (United States)},
  doi          = {10.2172/1340171},
  url          = {https://www.osti.gov/biblio/1340171},
  place        = {United States},
  year         = {2016},
  month        = {08}}

@article{barber2025,
  title = {Greater than 1000-Fold {{Gain}} in a {{Free-Electron Laser Driven}} by a {{Laser-Plasma Accelerator}} with {{High Reliability}}},
  author = {Barber, S. K. and Kohrell, F. and Doss, C. E. and Jensen, K. and Berger, C. and Isono, F. and Eisentraut, Z. and Schr{\"o}der, S. and Gonsalves, A. J. and Nakamura, K. and Plateau, G. R. and {van Mourik}, R. A. and {Gracia-Linares}, M. and Labun, L. and Hegelich, B. M. and Milton, S. V. and Geddes, C. G. R. and Osterhoff, J. and Schroeder, C. B. and Esarey, E. H. and {van Tilborg}, J.},
  year = {2025},
  month = jul,
  journal = {Physical Review Letters},
  volume = {135},
  number = {5},
  pages = {055001},
  publisher = {American Physical Society},
  doi = {10.1103/vh62-gz1p}
}

@article{zeng2014,
  title = {Self-Truncated Ionization Injection and Consequent Monoenergetic Electron Bunches in Laser Wakefield Acceleration},
  author = {Zeng, Ming and Chen, Min and Sheng, Zheng-Ming and Mori, Warren B. and Zhang, Jie},
  year = {2014},
  month = mar,
  journal = {Physics of Plasmas},
  volume = {21},
  number = {3},
  pages = {030701},
  issn = {1070-664X, 1089-7674},
  doi = {10.1063/1.4868404}
}

@article{mirzaie2015,
  title = {Demonstration of Self-Truncated Ionization Injection for {{GeV}} Electron Beams},
  author = {Mirzaie, M. and Li, S. and Zeng, M. and Hafz, N. A. M. and Chen, M. and Li, G. Y. and Zhu, Q. J. and Liao, H. and Sokollik, T. and Liu, F. and Ma, Y. Y. and Chen, L.M. and Sheng, Z. M. and Zhang, J.},
  year = {2015},
  month = oct,
  journal = {Scientific Reports},
  volume = {5},
  number = {1},
  pages = {14659},
  issn = {2045-2322},
  doi = {10.1038/srep14659}
}

@article{irman2018,
  title = {Improved Performance of Laser Wakefield Acceleration by Tailored Self-Truncated Ionization Injection},
  author = {Irman, A. and Couperus, J. P. and Debus, A. and K{\"o}hler, A. and Kr{\"a}mer, J. M. and Pausch, R. and Zarini, O. and Schramm, U.},
  year = {2018},
  month = mar,
  journal = {Plasma Physics and Controlled Fusion},
  volume = {60},
  number = {4},
  pages = {044015},
  publisher = {IOP Publishing},
  issn = {0741-3335},
  doi = {10.1088/1361-6587/aaaef1}
}

@misc{lindstrom2025,
  title = {Beam-Driven Plasma-Wakefield Acceleration},
  author = {Lindstr{\o}m, C. A. and Corde, S. and D'Arcy, R. and Gessner, S. and Gilljohann, M. and Hogan, M. J. and Osterhoff, J.},
  year = {2025},
  month = apr,
  number = {arXiv:2504.05558},
  eprint = {2504.05558},
  primaryclass = {physics},
  publisher = {arXiv},
  doi = {10.48550/arXiv.2504.05558}
}

@article{lotov2005,
  title = {Efficient Operating Mode of the Plasma Wakefield Accelerator},
  author = {Lotov, K. V.},
  year = {2005},
  month = may,
  journal = {Physics of Plasmas},
  volume = {12},
  number = {5},
  pages = {053105},
  issn = {1070-664X, 1089-7674},
  doi = {10.1063/1.1889444}
}

@article{chen1986,
  title = {Energy {{Transfer}} in the {{Plasma Wake-Field Accelerator}}},
  author = {Chen, Pisin and Su, J. J. and Dawson, J. M. and Bane, K. L. F. and Wilson, P. B.},
  year = {1986},
  month = mar,
  journal = {Physical Review Letters},
  volume = {56},
  number = {12},
  pages = {1252--1255},
  issn = {0031-9007},
  doi = {10.1103/PhysRevLett.56.1252}
}

@inproceedings{bane1985,
  title = {Wake Fields and Wake Field Acceleration},
  booktitle = {{{AIP Conference Proceedings}}},
  author = {Bane, K. L. F. and Wilson, P. B. and Weiland, T.},
  year = {1985},
  volume = {127},
  pages = {875--928},
  publisher = {AIP},
  issn = {0094243X},
  doi = {10.1063/1.35182},
  urldate = {2025-10-24}
}

@article{emma2025,
  title = {Experimental {{Generation}} of {{Extreme Electron Beams}} for {{Advanced Accelerator Applications}}},
  author = {Emma, C. and Majernik, N. and Swanson, K. K. and Ariniello, R. and Gessner, S. and Hessami, R. and Hogan, M. J. and Knetsch, A. and Larsen, K. A. and Marinelli, A. and O'Shea, B. and Perez, S. and Rajkovic, I. and Robles, R. and Storey, D. and Yocky, G.},
  year = {2025},
  month = feb,
  journal = {Physical Review Letters},
  volume = {134},
  number = {8},
  pages = {085001},
  publisher = {American Physical Society},
  doi = {10.1103/PhysRevLett.134.085001}
}

@article{oumbarekespinos2023,
  title = {Notable Improvements on {{LWFA}} through Precise Laser Wavefront Tuning},
  author = {Oumbarek Espinos, Driss and Rondepierre, Alexandre and Zhidkov, Alexei and Pathak, Naveen and Jin, Zhan and Huang, Kai and Nakanii, Nobuhiko and Daito, Izuru and Kando, Masaki and Hosokai, Tomonao},
  year = {2023},
  month = oct,
  journal = {Scientific Reports},
  volume = {13},
  number = {1},
  pages = {18466},
  publisher = {Nature Publishing Group},
  issn = {2045-2322},
  doi = {10.1038/s41598-023-45737-5}
}

@article{hue2023a,
  title = {Beam Current from Downramp Injection in Electron-Driven Plasma Wakefields},
  author = {Hue, C{\'e}line and Golovanov, Anton and Tata, Sheroy and Corde, S{\'e}bastien and Malka, Victor},
  year = {2023},
  month = oct,
  journal = {Journal of Plasma Physics},
  volume = {89},
  number = {5},
  pages = {965890502},
  issn = {0022-3778, 1469-7807},
  doi = {10.1017/S0022377823001162}
}

@article{zhang2025,
  title = {Plasma-Wakefield Accelerator Simultaneously Boosts Electron Beam Energy and Brightness},
  author = {Zhang, Chaojie and Storey, Douglas and Knetsch, Alexander and O'Shea, Brendan D. and Ariniello, Robert and Cao, Gevy J. and Corde, S{\'e}bastien and Dalichaouch, Thamine N. and Emma, Claudio and Finnerud, Ole G. and Gessner, Spencer and Hansel, Claire and Hansen, Elias and Lee, Valentina and Lindstr{\o}m, Carl A. and Litos, Michael and Majernik, Nathan and Marsh, Kenneth A. and Mori, Warren B. and Rajkovic, Ivan and Hogan, Mark J. and Joshi, Chan},
  year = {2025},
  month = nov,
  journal = {Nature Communications},
  volume = {16},
  number = {1},
  pages = {10719},
  issn = {2041-1723},
  doi = {10.1038/s41467-025-65742-8}
}

@article{salgado2021,
  title = {Towards Pair Production in the Non-Perturbative Regime},
  author = {Salgado, F C and Grafenstein, K and Golub, A and D{\"o}pp, A and Eckey, A and Hollatz, D and M{\"u}ller, C and Seidel, A and Seipt, D and Karsch, S and Zepf, M},
  year = {2021},
  month = oct,
  journal = {New Journal of Physics},
  volume = {23},
  number = {10},
  pages = {105002},
  issn = {1367-2630},
  doi = {10.1088/1367-2630/ac2921}
}

@inproceedings{habib2019,
  title = {Plasma Accelerator-Based Ultrabright x-Ray Beams from Ultrabright Electron Beams},
  booktitle = {Advances in {{Laboratory-based X-Ray Sources}}, {{Optics}}, and {{Applications VII}}},
  author = {Habib, Fahim and Scherkl, Paul and Manahan, Grace Gloria and Heinemann, Thomas and Ullmann, Daniel and Sutherland, Andrew and Knetsch, Alexander and Litos, Michael and Hogan, Mark and Rosenzweig, James and Hidding, Bernhard},
  editor = {Murokh, Alex and Spiga, Daniele},
  year = {2019},
  month = sep,
  pages = {9},
  publisher = {SPIE},
  address = {San Diego, United States},
  doi = {10.1117/12.2530976}
}

@article{habib2023,
  title = {Attosecond-{{Angstrom}} Free-Electron-Laser towards the Cold Beam Limit},
  author = {Habib, A. F. and Manahan, G. G. and Scherkl, P. and Heinemann, T. and Sutherland, A. and Altuiri, R. and Alotaibi, B. M. and Litos, M. and Cary, J. and Raubenheimer, T. and Hemsing, E. and Hogan, M. J. and Rosenzweig, J. B. and Williams, P. H. and McNeil, B. W. J. and Hidding, B.},
  year = {2023},
  month = feb,
  journal = {Nature Communications},
  volume = {14},
  number = {1},
  pages = {1054},
  publisher = {Nature Publishing Group},
  issn = {2041-1723},
  doi = {10.1038/s41467-023-36592-z}
}

@article{wood2026,
  title = {Bright Electron Bunches from a Plasma-Wakefield Accelerator with a Steep Density down-Ramp},
  author = {Wood, J. C. and Boulton, L. and Beinortait{\.e}, J. and Bj{\"o}rklund Svensson, J. and Bohlen, S. and Boyle, G. and Garland, J. M. and Gonzalez Caminal, P. and Lindstr{\o}m, C. A. and Loisch, G. and Mewes, S. M. and Parikh, T. and Pe{\~n}a, F. and P{\~o}der, K. and Schr{\"o}der, S. and Th{\'e}venet, M. and Wesch, S. and Osterhoff, J. and D'Arcy, R.},
  year = {2026},
  month = feb,
  journal = {Nature Communications},
  volume = {17},
  number = {1},
  pages = {1588},
  issn = {2041-1723},
  doi = {10.1038/s41467-026-69283-6}
}

@inproceedings{wood:ipac2024-mopr40,
    author = {J. Wood et al.},
    title = {Progress towards high-quality, high-repetition-rate plasma acceleration at FLASHForward},
    booktitle = {Proc. IPAC'24},
    pages = {541-544},
    paper = {MOPR40},
    venue = {Nashville, TN},
    series = {IPAC'24 - 15th International Particle Accelerator Conference},
    number = {15},
    publisher = {JACoW Publishing, Geneva, Switzerland},
    month = {05},
    year = {2024},
    issn = {2673-5490},
    isbn = {978-3-95450-247-9},
    doi = {10.18429/JACoW-IPAC2024-MOPR40},
    url = {https://indico.jacow.org/event/63/contributions/3551}
}
\end{document}



\title{Supplemental Material: Efficient Acceleration of High-Quality GeV-Electron Bunches in a Hybrid Laser- and Beam-Driven Plasma Wakefield Accelerator}

\author{F. M. Foerster}
\email{moritz.foerster@physik.uni-muenchen.de}
\affiliation{Ludwig--Maximilians--Universit{\"a}t M{\"u}nchen, Am Coulombwall 1, 85748 Garching, Germany}%

\author{M. Ayache}
\affiliation{Ludwig--Maximilians--Universit{\"a}t M{\"u}nchen, Am Coulombwall 1, 85748 Garching, Germany}%

\author{Z. Bi}
\affiliation{Ludwig--Maximilians--Universit{\"a}t M{\"u}nchen, Am Coulombwall 1, 85748 Garching, Germany}%

\author{M. Cerchez}
\affiliation{Heinrich Heine Universität Düsseldorf, Universitätsstraße 1, 40225 Düsseldorf, Germany}%

\author{S. Corde}
\affiliation{Laboratoire d’Optique Appliquée (LOA), CNRS, École polytechnique, ENSTA, Institut Polytechnique de Paris, Palaiseau, France}%

\author{A. D{\"o}pp}
\affiliation{Ludwig--Maximilians--Universit{\"a}t M{\"u}nchen, Am Coulombwall 1, 85748 Garching, Germany}%
\affiliation{Max Planck Institut für Quantenoptik, Hans-Kopfermann-Strasse 1, 85748 Garching , Germany}%

\author{F. Haberstroh}
\affiliation{Ludwig--Maximilians--Universit{\"a}t M{\"u}nchen, Am Coulombwall 1, 85748 Garching, Germany}%

\author{A. F. Habib}
\affiliation{University of Strathclyde, 107 Rottenrow, Glasgow G4 0NG, United Kingdom}

\author{T. Heinemann}
\affiliation{Heinrich Heine Universität Düsseldorf, Universitätsstraße 1, 40225 Düsseldorf, Germany}%

\author{B. Hidding}
\affiliation{Heinrich Heine Universität Düsseldorf, Universitätsstraße 1, 40225 Düsseldorf, Germany}%

\author{A. Irman}
\affiliation{Helmholtz-Zentrum Dresden--Rossendorf, Bautzner Landstrasse 400, 01328 Dresden, Germany}%

\author{F. Irshad}
\affiliation{Ludwig--Maximilians--Universit{\"a}t M{\"u}nchen, Am Coulombwall 1, 85748 Garching, Germany}%

\author{O. Kononenko}
\affiliation{Laboratoire d’Optique Appliquée (LOA), CNRS, École polytechnique, ENSTA, Institut Polytechnique de Paris, Palaiseau, France}%

\author{M. LaBerge}
\affiliation{Helmholtz-Zentrum Dresden--Rossendorf, Bautzner Landstrasse 400, 01328 Dresden, Germany}%

\author{A. Martinez de la Ossa}
\affiliation{Deutsches Elektronen-Synchrotron DESY, Notkestraße 85, 22607 Hamburg, Germany}%

\author{A. Münzer}
\affiliation{Ludwig--Maximilians--Universit{\"a}t M{\"u}nchen, Am Coulombwall 1, 85748 Garching, Germany}%

\author{F. Pe\~na}
\affiliation{Ludwig--Maximilians--Universit{\"a}t M{\"u}nchen, Am Coulombwall 1, 85748 Garching, Germany}%
\affiliation{Department of Physics, University of Oslo, 0316, Oslo, Norway}

\author{G. Schilling}
\affiliation{Ludwig--Maximilians--Universit{\"a}t M{\"u}nchen, Am Coulombwall 1, 85748 Garching, Germany}%

\author{S. Sch{\"o}bel}
\affiliation{Helmholtz-Zentrum Dresden--Rossendorf, Bautzner Landstrasse 400, 01328 Dresden, Germany}%
\affiliation{Technische Universität Dresden, 01062 Dresden, Germany}

\author{U. Schramm}
\affiliation{Helmholtz-Zentrum Dresden--Rossendorf, Bautzner Landstrasse 400, 01328 Dresden, Germany}
\affiliation{Technische Universität Dresden, 01062 Dresden, Germany}%

\author{S. Sharan}
\affiliation{Ludwig--Maximilians--Universit{\"a}t M{\"u}nchen, Am Coulombwall 1, 85748 Garching, Germany}%

\author{E. Travac}
\affiliation{Ludwig--Maximilians--Universit{\"a}t M{\"u}nchen, Am Coulombwall 1, 85748 Garching, Germany}%

\author{P. Ufer}
\affiliation{Helmholtz-Zentrum Dresden--Rossendorf, Bautzner Landstrasse 400, 01328 Dresden, Germany}%
\affiliation{Technische Universität Dresden, 01062 Dresden, Germany}

\author{N. Weiße}
\affiliation{Ludwig--Maximilians--Universit{\"a}t M{\"u}nchen, Am Coulombwall 1, 85748 Garching, Germany}%

\author{M. Zeuner}
\affiliation{Ludwig--Maximilians--Universit{\"a}t M{\"u}nchen, Am Coulombwall 1, 85748 Garching, Germany}%

\author{J. Zirkelbach}
\affiliation{Ludwig--Maximilians--Universit{\"a}t M{\"u}nchen, Am Coulombwall 1, 85748 Garching, Germany}%

\author{S. Karsch}
\email{stefan.karsch@physik.uni-muenchen.de}
\affiliation{Ludwig--Maximilians--Universit{\"a}t M{\"u}nchen, Am Coulombwall 1, 85748 Garching, Germany}%
\affiliation{Max Planck Institut für Quantenoptik, Hans-Kopfermann-Strasse 1, 85748 Garching , Germany}%





\date{\today}

\maketitle

\section{Analysis of driver-to-witness efficiency}
To quantify the energy transfer efficiency from driver to witness we use the following definition in our analysis:

\begin{equation*}
    \eta_{D\rightarrow W} = \frac{\Delta \varepsilon_{\text{W}}}{\varepsilon_{\text{D}}}. 
\end{equation*}

The calculated efficiency values are subject to different types of uncertainties. 
The integrated energy of driver and witness $\varepsilon_{\text{D/W}} = \int \rho_Q\left(E\right) E dE$ fluctuates due to shot-to-shot variations of the spectral charge density $\rho_Q$, i.e. of bunch charge and peak energy of both bunches. 
Examining set averages, these shot-to-shot variations are by far the largest contributors to the uncertainty interval. For the drive bunches, the relative standard deviation of the integrated energy is approximately 20\%. For the witness bunch, the value of these fluctuations depends on the injection position. For the highest-energy witness bunches, those injected at the very front of the PWFA target, the shot-to-shot fluctuations peak at a relative standard deviation of approximately 70\%.

Furthermore, individual shots are subject to measurement uncertainties, which, however, are significantly lower than the shot-to-shot fluctuations:

The absolute charge calibration of our dipole spectrometer does not pose a major problem, as the charge calibration constant cancels out by dividing the integrated energies of both bunches. 

The energy calibration of the spectrometer is an issue. In particular, the measured energy depends on the pointing of the electron beams due to the trajectory of the electron beams in the dipole spectrometer. No pointing screen was used for the data presented in this article, as this increases beam divergence; thus, beam quality and energy cannot be measured accurately simultaneously. 

Instead, we estimate the energy uncertainty by analyzing the pointing jitter in a different run under similar conditions. We measure shot-to-shot pointing fluctuations of $\SI{5}{mrad}$ (standard deviation), which corresponds to an energy error of $ \pm 5\%$ for our dipole spectrometer at a peak energy of $\SI{1}{GeV}$. Furthermore, we don't observe any significant correlation between pointing and peak energy of the witness, i.e.~the shot-to-shot fluctuations in energy are larger than the effect of varying pointing. 

A general problem with PWFA experiments is that we cannot measure drive bunch properties before and after the PWFA stage individually on a single shot. 

To account for the sum of these effects, we estimate a lower limit for the peak driver-to-witness energy transfer efficiency by dividing the integrated witness energy of the best single shot by the highest integrated driver energy measured among all analyzed LWFA shots.

\section{Comparison of PWFA results}
In Table~\ref{Table:efficiency}, we present an overview of driver and witness parameters, transformer ratio, and energy transfer efficiencies reported from PWFA experiments. The data from this table have been used to compile Figure~4 of the manuscript. We would like to point out that not all authors provide exactly identical definitions of the beam parameters cited here. Some of the values were taken only approximately from the authors’ figures. However, this does not fundamentally affect their comparability. The values for blank spaces in the table could not be determined from or do not apply to the respective publications.

\begin{table*}
\caption{Overview of beam parameters, transformer ratio, and energy transfer efficiencies reported from PWFA experiments. Witness and driver charges are denoted by $Q_{W}$ and $Q_{D}$. $\Delta E_{W}$ is the witness energy gain and $E_{D}$ the initial driver energy. $\frac{\delta E_{W}}{E_{W}}$ is the relative energy spread of the witness bunch. $\eta_{D\rightarrow W}$ is the overall driver-to-witness energy transfer efficience, and $\eta_{P\rightarrow W}$ is the plasma-to-witness energy transfer efficiency.}
\begin{tabular}{l|lll|llll|ll|l}
                             & Driver & $Q_{D}$  & $E_{D}$  & Inj. & $Q_{W}$  & $\Delta E_{W}$ & $\frac{\delta E_{W}}{E_{W}}$ & $\eta_{D\rightarrow W}$ & $\eta_{P\rightarrow W}$  & $\frac{\Delta E_{W}}{E_{D}}$ \\
                            Publication &   &   [pC] &   [MeV] &   &   [pC] &   [MeV] &   [\%] &   [\%] &   [\%] &   [\%] \\
\hline
\hline
                   This work, Fig. 2(c) &       LWFA &       366 &        716 & int. &        48 &         930 &    4.1 (FWHM)       &     $\geq 9.5$ &      $\geq 21$ &               130 \\
                   This work, Fig. 2(d) &       LWFA &       366 &        716 & int. &        53 &         1080 &    6.9 (FWHM)       &     $\geq17$ &      $\geq39$ &               150 \\
                   This work, Fig. 2(e) &       LWFA &       366 &        716 & int. &        38 &         1280 &    3.1 (FWHM)       &     $\geq 7.0$ &      $\geq 16$ &               180 \\
\hline
Blumenfeld2007~\cite{Blumenfeld:2007ja} &      rf &           &      42000 & ext. &           &       44000 &                &        &         &               105 \\
          Litos2014~\cite{Litos:2015ke} &      rf &      1020 &      20350 & ext. &        74 &        1600 &                 2 (rms) &    0.6 &      29 &                 8 \\
            Litos2016~\cite{Litos:2016} &      rf &       600 &      20350 & ext. &       120 &        5300 &                 5 (rms) &    5.2 &         &                26 \\
                  Vafaei-Najafabadi2016~\cite{vafaei-najafabadi2016} &      rf &    2880   &   20350    & int. &      25   &   30000      &  20  (FWHM)           &   1.3 &         &   150              \\
                             Loisch2018~\cite{loisch2018} &      rf &    500    &  25        & ext. &     3    &  0.4    &               &    0.01    &         &  1.6              \\
                               Deng2019~\cite{deng2019} &      rf &  3200     &  20350     & int. &      100  &    500      &   2.5  (rms)           &  0.1   &         &    2.5            \\
                            Roussel2020~\cite{roussel2020} &      rf &   1600    &   40      & ext. &           &      1      &                   &        &         &      2.5          \\
  Lindstrom2021~\cite{Lindstrom:2021eo} &      rf &       500 &       1000 & ext. &       100 &          45 &              0.13 (FWHM)    &    0.9 &      42 &                 5 \\
      Knetsch2021~\cite{Knetsch:2021kj} &      rf &       790 &       1116 & int. &        33 &          45 &         4.4 (rms)          &    0.2 &         &                 4 \\
      Ullmann2021~\cite{Ullmann:2021ky} &      rf &      3200 &      20000 & int. &       500 &         780 &          6.4 (rms)       &    0.6 &         &                 4 \\
            Kurz2021~\cite{Kurz:2021cg} &       LWFA &       104 &        260 & ext. &        12 &          66 &        6 (FWHM)       &    2.9 &         &                25 \\
         Pompili2021~\cite{pompili2021} &      rf &       350 &       89.5 & ext. &        20 &         6.8 &       0.1 (rms)          &    0.4 &         &                 8 \\
         Foerster2022~\cite{foerster2022} &      LWFA &       650  &  250   & int. &   30    &   162    &    3.5 (FWHM)              &  3   &         &                 65 \\
             Lindstrom2024~\cite{lindstrom2024} &      rf &       400 &       1050 & ext. &        40 &          40 &              0.09 (FWHM) &    0.4 &      22 &                 4 \\
             Wood2024~\cite{wood:ipac2024-mopr40} &      rf &       230 &       1200 & ext. &        40 &          250 &              0.17 (FWHM) &    3.6 &        &                 21 \\
                              Zhang2024~\cite{zhang2024} &      rf &    1600   &   10000   & ext. &           &   8000     &               &        &         &      80             \\
            Zhang2025~\cite{zhang2025} &      rf &    1600   &   10000   & int. &   2.3     &   26000     &   0.7 (rms)            &   0.4     &         &      260             \\
             Wood2026~\cite{wood2026} &      rf &       304 &       689 & int. &        19 &          30 &              1.3 (FWHM) &    0.3 &        &                 4.4 \\
\end{tabular}
\label{Table:efficiency}
\end{table*}

\section{Particle in cell simulations}
This paragraph provides information on the input of the PIC simulations of the PWFA stage used in Figure~1 of the manuscript. The physical parameters are similar to our experiments and the results reproduce our experiment qualitatively.

We use the PIC-code FBPIC (Fourier-Bessel Particle-In-Cell)~\cite{FBPIC,Lehe:2016dn}. 
We make use of the cylindrical coordinate system used in FBPIC, by assuming complete rotational symmetry and, therefore, use only one azimuthal mode. To resolve the injection of the witness bunch at the density transition, we use a spatial resolution of $\Delta z = \SI{100}{nm}$ and $\Delta r = \SI{350}{nm}$. Furthermore, we increase the number of macroparticles per cell around the density transition to $n_{z} = 10$ (Number of particles per cell along z), $n_{r} = 4$ (along r), and $n_{t} = 1$ (along theta), while it is 2, 2, and 1, respectively, elsewhere. 

The plasma density profile is presented in Figure~1 of the manuscript and models the one we retrieved from interferometry of the plasma. 

\begin{figure}[t]
  \centering
  \includegraphics*[width=\linewidth]{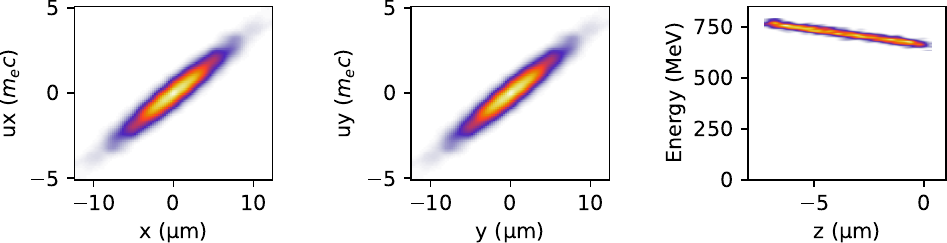}
  \caption{Transverse and longitudinal phase space distributions of the simulated drive bunch just before entering the PWFA stage.}
  \label{fig:phase_space}
\end{figure}

We initialize the phase space of the drive bunch, approximating the experimental measurements. The driver's mean energy is $\SI{700}{MeV}$, with an uncorrelated energy spread of $\SI{10}{MeV}$ (rms), and a linear chirp leading to a correlated energy spread of $\SI{200}{MeV}$. The bunch length is $\SI{7}{\micro m}$. 
Source size and divergence of the driver are set to $\SI{1}{\micro m}$ (rms) and $\SI{1}{mrad}$ (rms), respectively. 
The phase space of the drive bunch just before entering the PWFA target is shown in Figure~\ref{fig:phase_space}.

\bibliography{references}